# Recommender Systems using Pennant Diagrams in Digital Libraries

Zeljko Carevic[1], Philipp Mayr

**Introduction**
Recommender systems in search systems are an established way of pointing the user to related content. Commercial companies like Amazon have been using recommendations for a while by showing the user products related to their current search context or usage behaviour. In digital libraries recommendations can be valuable for researchers, e.g. recommending related literature to a given context. Typically, in a scientific context the simple presentation of related content is not sufficient. Often the users demand a more detailed view on the connection of a document and its specific recommendations. Key questions are: What is mostly related to the current document? Who are the main researchers in a particular area? Which research topics are related to the current document? One way of visualizing the relatedness / distance between a given document and it's recommendations are the so-called pennant diagrams. The aim of pennants is to provide the user with a graph showing the relatedness / distance between a given document and related documents. Co-citation but also co-occurrence analysis are established methods for finding related documents to a seed. A seed could be for instance an author, a keyword, or a publication. However, these approaches are not suitable to make any assumption about the distance between the seed and its co-mentioned documents.
In this paper we introduce a recommender system in the digital library sowiport[2] using pennant diagrams which can be created from co-citation and/or co-occurrence analysis.

**Background**
In (2007a, b, 2009, 2010), White proposes a new method for visualizing co-mentioned documents called pennant diagrams. Using methods from information retrieval, bibliometrics, and relevance theory, pennant diagrams can be used to visualize the distance of a given seed to its co-mentioned documents. When creating pennant diagrams, two criteria from relevance theory are being used: (i) cognitive effect and (ii) processing effort. Regarding (i), the closer a co-mentioned document is to the seed the higher the cognitive effect. Considering (ii) on the other hand, the less cognitive effort is necessary to connect the seed and the co-mentioned document, the lower is the processing effort. Pennant diagrams are divided into three sectors depicting the degree of specificity. Co-mentioned documents in sector A can be considered as "see also" references. Co-mentioned documents within the sectors B and C are broader related to seed. These sectors can be of value within a literature search but do not apparently relate to the seed.

---

[1] Corresponding author
[2] http://sowiport.gesis.org/ - portal for the social sciences.

White & Mayr (2013) have proposed a method how to apply pennants in an environment where co-citations are missing but where co-occurrences of descriptors are used to visualize pennants. They showed that a pennant diagram can be constructed only using descriptors co-mentioned in a database like sowiport.

**Use Case**

To the best of our knowledge, there is no digital library using pennant diagrams. Two possible use cases for pennant diagrams in digital libraries could be as follows: a) a recommender tool showing related documents and b) a tool to explore the information space (e.g. descriptors) within digital libraries like sowiport. A typically use case could be a researcher looking for documents broader related to a given topic. By dividing the pennant into three sectors the user is able to get an overview of the distance between a seed and the recommended documents. Another use case could be a researcher looking for topics related to a given descriptor. Again the pennant diagram shows related descriptors and their distances to the seed term.

We will implement pennant diagrams to display related documents and their distance to a given seed. In this case, the seed is instantiated by the user looking at a specific document within sowiport. Subsequently, the system provides the user with a pennant diagram. When creating pennant diagrams for co-citations we first have to identify references for the given seed. The identification of references is done by using sowiport's reference database containing about 8.0 million cited references (Sawitzki et al., 2013). Calculating the cognitive effect and the processing effort is done by using the well-known tf*idf formula. For calculating the tf*idf measure we applied co-citation analysis to the seed document where tf is the citation count of the seed document together with a potential candidate at the same time. Correspondingly df is the overall citation count for the candidate in the whole reference corpus (Carevic and Schaer, 2014). The logged frequencies of the co-citation count is construed as the cognitive effect and the logged inverse count of the overall citation is construed as the predicted processing effort. Calculating tf and idf from co-occurring descriptors is done by counting how often a descriptor co-occurs with the seed (tf) and how often a descriptor occurs in the database (df).

The presentation at the NKOS workshop will present demos of pennants in sowiport and will elaborate on practical questions in visualizing pennants and evaluating the utility of pennants for search.